\documentclass[12pt,english]{article}
\usepackage[T1]{fontenc}
\usepackage{geometry}
\usepackage{float}
\usepackage{amsmath}
\usepackage{amssymb}
\usepackage{graphicx}
\usepackage{setspace,color}  
\usepackage[authoryear]{natbib}
\usepackage[position=top]{subfig}
\usepackage{url}   
\usepackage{hyperref} 
\usepackage{xcolor}  
\hypersetup{
  colorlinks   = true, 
  urlcolor     = blue, 
  linkcolor    = blue, 
  citecolor   = black 
}
\usepackage[labelfont={bf},textfont={}]{caption} 

\makeatletter 

\def \E{\mathbb{E}}
\title{A Spatial Stochastic SIR Model for Transmission Networks with Application to COVID-19 Epidemic in China\thanks{
  We would
  like to acknowledge the financial support from
  the Centre for Development Economics and Sustainability (CDES) at Monash University.
  Yunyun Wang provided superb research assistance.}
}

\author{
Tatsushi Oka\thanks{\noindent Department of Econometrics and Business Statistics, 	
  Monash University
  (\texttt{tatsushi.oka@monash.edu}).}
\hspace{2cm}
Wei Wei\thanks{\noindent Department of Econometrics and Business Statistics, 	
Monash University 
(\texttt{wei.wei2@monash.edu}).}
\hspace{2cm}
Dan Zhu\thanks{\noindent Department of Econometrics and Business Statistics, 	
Monash University 
(\texttt{dan.zhu@monash.edu}).}
}

\usepackage{babel}

\@ifundefined{showcaptionsetup}{}{%
 \PassOptionsToPackage{caption=false}{subfig}}
\usepackage{subfig}
\makeatother

\usepackage{babel}
\begin{document}
\maketitle 
\begin{abstract}

Governments around the world have implemented preventive measures against the spread of the coronavirus disease (COVID-19). In this study, we consider a multivariate discrete-time Markov model to analyze the propagation of COVID-19 across 33 provincial regions in China. This approach enables us to evaluate the effect of mobility restriction policies on the spread of the disease. We use data on daily human mobility across regions and apply the Bayesian framework to estimate the proposed model. The results show that the spread of the disease in China was predominately driven by community transmission within regions and the lockdown policy introduced by local governments curbed the spread of the pandemic. Further, we document that Hubei was only the epicenter of the early epidemic stage. Secondary epicenters, such as Beijing and Guangdong, had already become established by late January 2020, and the disease spread out to connected regions. The transmission from these epicenters substantially declined following the introduction of human mobility restrictions across regions. 
\end{abstract}

\vspace{0.5cm}

\newpage
\section{Introduction}

The ongoing pandemic of coronavirus disease (COVID-19) poses a threat to public health and has disrupted economic activities globally. Although there are limited policy tools available to stem the disease spread, restricting human mobility though lockdown or border closure policies was identified as an effective measure. Simply put, the virus itself cannot move anywhere without assistance. In many countries, mobility restriction led to the containment of the virus's spread. Given the importance of mobility restriction as an effective policy, it is critical to quantify its effects.

In this study, we consider a multivariate discrete-time Markov model to analyze the propagation of COVID-19 across 33 provincial regions of China. Thereby, we allow for heterogeneous disease transmission both within and across regions.\footnote{
  China consists of 
  27 provinces, four special administrative cities in mainland, and two special administrative regions (Hong Kong and Macao). Through the paper, we use  ``region'' for the provinces and special administrative regions.} 
Our model takes into account human mobility as a key driver of disease transmission across regions and identifies epicenters of disease propagation, as well as the effect of mobility restrictions on infection rates. We extract information on daily human mobility across regions from January 11 to March 15, 2020, from the
\cite{Baidu} database and apply the Bayesian framework to estimate the model. The sampling period in use for our analysis exhibits substantial exogenous variations in human mobility rates due to the high number of movements around Chinese New Year (January 25) and a sudden decline in movements after policy interventions were introduced. We evaluate the effect of mobility restrictions on the disease spread between regions by comparing outcomes under actual and counterfactual human mobility, which is extracted from the 2019 data.

Our empirical results document substantial heterogeneity in the rate of infection across regions.  The results also demonstrate the effectiveness of the lockdown policy in curbing the spread of the pandemic.  The transmission mechanism of the disease in China is found to be predominately community transmission within all regions. Further, our analysis based on the 2019 mobility data suggests that the external transmission would not have been suppressed if people had continued to be allowed to move freely across provincial borders as usual.  Interestingly, our results show that Hubei is not the only epicenter of the early epidemic stage. Other epicenters, such as Beijing and Guangdong, had already become established by late January 2020.
The pandemic radiated out to the subordinate regions of these cities with varying degrees of severity. Our approach sheds light on the evolution of the transmission network over time and provides useful insight into the formulation of lockdown policies amid the pandemic.

The methodological part of the paper draws on and contributes to several literatures.
First, since the outbreak of COVID-19, many studies have provided simulations and predictions using a deterministic susceptible-infective-recovered (SIR) model
\citep{kermack1927contribution}. 
The SIR model divides a well-defined population into three compartments,
namely susceptible, infective, and recovered individuals,
and characterizes 
disease transmission as individuals' transition between these compartments
\citep[e.g.,][]{held2019handbook}.
As \cite{allen2017primer} discusses, however, stochastic modeling of
epidemics is essential when the number of infectious individuals is small, and the transition between the compartments depends on demography and the environment.
We consider a variation of a stochastic SIR model for the 33 regions in China.

Second, the most critical feature of our model is that it captures the impact of human
movements on spatio-temporal disease transmission.
The quantitative modeling of human movements has a long-standing history
in fields like transportation, tourism, and urban planning.
The use of the gravity model has been popular in these fields 
\citep{wilson1974urban, haynes1984gravity, erlander1990gravity}
as well as in the field of economics
\citep{anderson2011gravity, head2014gravity}.
In the epidemiology literature,
the gravity model was first applied by \cite{murray1977stochastic}.
Gravity-type models 
are also widely adopted in more recent studies 
\citep{xia2004measles, viboud2006synchrony, balcan2009multiscale, jandarov2014emulating}.
Alternatively, the radiation model, proposed by \cite{simini2012universal},
is used  to predict spatial disease transmission
\citep{tizzoni2014use, kraemer2019utilizing}.
Both the gravity and radiation models treat the transition probabilities of individuals from one place to another as a function of population sizes and geometric distances, both of which are almost invariant on a daily basis.
By contrast, this study uses known information on daily human mobility to characterize disease transmission across regions and evaluate the dynamic impact of mobility restrictions.

Third, there is a growing body of literature dedicated to the study of the spread of infections in China.
\cite{wu2020nowcasting}
study the transmission of COVID-19 from Wuhan to other cities, combining monthly and daily human mobility data up to January 28, 2020.\footnote{
  They combine three data sets:
  1)
  the monthly number of domestic and international flight bookings from Wuhan in January to February  2019,
  2)
  the number of daily domestic passengers by train and car,
  and
  3)
  travel volumes forecast from and to Wuhan
  by Wuhan Municipal Transportation Management Bureau.}
\cite{kraemer2020effect}
use human mobility information from Baidu-Qianxi and analyze the disease spread from Wuhan to other regions between January 1 and February 10, 2020. They predict daily case counts in the early phase of disease spread using three different models: Poisson, negative binomial, and log-linear regression.
Both \cite{wu2020nowcasting} and \cite{kraemer2020effect}
document the significance of human mobility from Wuhan in causing the spread of the disease in the early phase. Both authors also underscore that the effect travel restrictions in Hubei had on containing the spread of the disease. In our study, we estimate a model that accounts for disease transmission across all regions, using data spanning from the beginning of the epidemic until the end of the first wave in China. Our result is consistent with the existing findings, in that we show that Hubei is the earliest epicenter. However, we also show that additional regions became epicenters in an early phase of the pandemic in China. Thus, our results suggest that local government interventions, such as lockdown in Wuhan, cannot fully explain the containment of the disease. Mobility restriction across regions is essential. Our research complements the existing research by providing a more complete understanding of the spread of the disease using a broad and well-defined framework.

Lastly, we contribute to the large body of literature
that analyzes infection control
in epidemiology
\citep{wickwire1977mathematical, sethi1978optimal}
and in economics
\citep{wiemer1987optimal, philipson2000economic, gersovitz2004economical, rowthorn2012optimal}.
The economic literature theoretically analyzes the optimal control of infection from the perspective of a social planner and discusses how public policies, such as subsidies or taxes, can provide individuals with the required incentives to achieve the social planner's first-best solution. Our empirical results underscore the importance of coordination between central and local governments. In the presence of human mobility across regions, regional epicenters can quickly become established and transmit the disease to connected regions. Thus, if the primary goal is to eliminate the disease entirely, the central and local governments must implement preventive measures simultaneously.

The remainder of the paper proceeds as follows. Section 2 introduces a variation of the stochastic SIR model and details the model specifications. Section 3 explains the data in use and the estimation method. Section 4 presents the estimation results and Section 5 concludes.

\section{Model}

This section first introduces the variation of the susceptible-infective-recovered (SIR) model applied in this study. Subsequently, it explains the specification of internal and external disease transmission in the model.

\subsection{Stochastic SIR Model with Spatial Effects}

We apply a variation of stochastic SIR model to describe the evolution
of three variables: $S_{jt}$, $I_{jt}$ and $R_{jt}$, which denote
the number of susceptible, infective, and recovered individuals in
region $j$ at time $t$, respectively. Also, let $D_{jt}$ denote the cumulative
number of deaths by $t$ and let $N_{j}$ be the total population
in region $j$. Then, we have the following identity: $N_{j}=S_{jt}+I_{jt}+R_{jt}+D_{jt}.$
We observe regional panel data of $(I_{jt},R_{jt},D_{jt},N_{j})$
for region $j=1,\dots,J$ and time $t=0,\dots,T$ with $J$ and $T$
denoting the sample size of regions and time periods, respectively.
In what follows, we use $\mathcal{F}_{t}$
to denote the available information set at time $t$.

We denote by $\Delta_{j,t+1}^{I}$
the number of transitions from susceptible to infected states
in region $j$ at time $t+1$.
The number of newly infected individuals
$\Delta_{j,t+1}^{I}$
is assumed to be a random variable following 
the Poisson distribution conditional on $\mathcal{F}_{t}$,  with the conditional mean given by 
\begin{eqnarray}
\label{eq:expected}
  \E[\Delta_{j,t+1}^{I}|\mathcal{F}_{t}]
  =\Big(\beta_{jt}\frac{I_{jt}}{N_{j}}+\lambda_{jt}\Big)S_{jt}.
  \label{eq:mean}
\end{eqnarray}
At its core, the equation above follows the Bass model \citep{bass1969new}, which was originally proposed for describing the diffusion of new products. The key feature of the Bass model is that the acceptance of a new product is driven by either internal influences, such as contagious adopters to which other individuals are connected, or external influences, such as mass media or commercials. The distinction between internal and external influences is adopted by  \cite{fibich2016bass} in a deterministic SIR model. Similarly, we can interpret the term
$\beta_{jt}I_{jt}/N$ as region $j$'s internal infection rate, which depends on the proportion of infected individuals $I_{jt}/N$  and the internal transmission rate $\beta_{jt}$.
Further, we consider the term  $\lambda_{jt}$ as the external infection rate, which reflects the rate of infection attributable to transmission from outside of region $j$. If the border to region $j$ is closed, the external effect $\lambda_{jt}$ equals zero and the model becomes the standard stochastic SIR model
\citep[e.g.,][]{allen2008introduction}.

To describe the state transition from the infected state,
we use a Markov chain model in which
infected individuals either remain infected or
move to another state: recovery or death.
More specifically, 
let 
$\Delta_{j, t+1}^{R}:=R_{j,t+1}-R_{jt}$
and 
$\Delta_{j,t+1}^{D}:=D_{j,t+1}-D_{jt} $
be changes in the number of recoveries and deaths, respectively.
We assume that the transition probability from the infected state at time $t$ follows a multinomial distribution conditional on $\mathcal{F}_{t}$, satisfying that 
$\E[\Delta_{j, t+1}^{R}|\mathcal{F}_{t}] = \gamma I_{j,t}$
and
$\E[\Delta_{j,t+1}^{D}|\mathcal{F}_{t}] = \delta I_{j,t}$.
Here, the parameters $\gamma$ and $\delta$
are used to represent the recovery and death rates, respectively.
Given the stochastic transition among all states, 
the number of infected and susceptible individuals at time $t+1$ are given by the following state equations: 
\begin{eqnarray}
  I_{j,t+1}& = & I_{jt}
                 + \Delta_{j,t+1}^{I}
                 - \Delta_{j,t+1}^{R} - \Delta_{j,t+1}^{D},\label{eq:I}\\
S_{j,t+1} & = & S_{jt}-\Delta_{j,t+1}^{I}.\label{eq:S}
\end{eqnarray}

\subsection{Internal Transmission}

The internal transmission rate $\beta_{jt}$
measures to what extent contacts between
an infected individual and the susceptible population at time $t$
leads to the transmission of the pathogen.
Thus, it can be interpreted as the number of ``effective'' contacts.
We allow for $\beta_{jt}$ to vary per region and across time.
This is because
the contact frequency depends on region-specific characteristics, such as population density,
as well as time-varying factors, such as
policy intervention (e.g.~contact tracing and
forced quarantine) and 
behavior changes (e.g.~better hygiene practices and social distancing).
In China, almost all local governments declared the top-level state of emergency in the early phase of the pandemic (January 23-25, 2020), which effectively induced changes in individuals' behavior. Thus, we assume that intervention by local governments affects internal transmission gradually. Specifically, we consider the following specification:
\begin{equation}
  \log\beta_{jt}=\log\beta_{j,t-1}+\alpha_{j}X_{j,t-h},\label{eq:betat}
\end{equation}
where $X_{j,t-h}$ is an observed dummy variable taking the value of 1 if
the local government in region $j$ has
activated the top-level health emergency response
at time $t-h$
and 0 otherwise. This $X_{j,t-h}$ reflects the implementation of various intervention policies that we collectively call the \emph{lockdown policy}.
We consider a lag $h>0$ to account for lagged effects of the policy intervention
and set four days ($h=4$) for our estimation.\footnote{
  In the existing literature 
  the mean incubation period of COVID-19 is estimated
  as roughly 5 days
  \citep[see][among others]{li2020early, kraemer2020effect}.
}
The parameter $\alpha_{j}$ is allowed to be heterogeneous
across regions, reflecting different measures taken by local governments
and regional characteristics.
The time-varying parameter $\beta_{jt}$ in (\ref{eq:betat}) depends on 
the initial value $\beta_{j,0}$ and the response to the intervention $\alpha_{j}$.\footnote{
  Our specification allows $\beta_{jt}$ to approach zero in consideration of the draconian measures adopted in China and the suppression of the disease in the first wave. Alternatively, $\beta_{jt}$ could be set to approach a non-zero value as in 
\cite{FJ2020}. Their dynamics can be considered as a special case of the transfer function model in \cite{BoxTiao1975} for intervention analysis.}
We specify a hierarchical structure for the transmission parameters
across regions, by using a bivariate normal distribution: 
$(\log\beta_{j,0},\alpha_{j})' \sim N(\mu,\Sigma)$
with mean $\mu:=(\mu_{\beta}, \mu_{\alpha})'$ and variance matrix $\Sigma$.
Under this specification, the average of the internal transmission rate without any control is given by $\E[\beta_{j,0}]=\exp(\mu_{\beta}+1/2\Sigma_{11})$
with $\Sigma_{11}$ denoting the (1,1)-element of $\Sigma$,
while
the effect of intervention on average is given by $\E[\alpha_{j}]=\mu_{\alpha}$.

\subsection{External Transmission}

Using Baidu's daily mobility data, we construct a measure of the ``intensity'' of the disease transmission between regions.
The mobility data includes an outflux mobility index for all regions
and
details the proportion of travelers between regions.
We use $M_{kt}^{out}$ to denote the outflux mobility index in region $k$ at time $t$
and we use $P_{kjt}$ to represent the proportion of travelers from region $k$ to region $j$
at time $t$.
The mobility index $M_{kt}^{out}$ represents a relative strength measure of the outflux,
which is scaled by Baidu's proprietary method, rather than the numbers of outflux.
This index is comparable across regions and time. The change of $M_{kt}^{out}$ from its standard level reflects \emph{mobility restrictions}, which we discuss in the supplementary material. 
Additionally, we observe the proportion of daily travelers $P_{kjt}$ between
the 31 mainland regions in the sample of 33 provincial regions, which means Hong Kong and Macao are excluded. We impute the entries for Hong Kong and Macao based on the radiation model \citep{simini2012universal}.
In our supplementary material, we demonstrate that the prediction of influx based on the imputed $P_{kjt}$ value traces the index of human influx well. It also outperforms the prediction using only the radiation model.

We use $M_{kt}^{out}P_{kjt}$ to measure the (scaled) flux from origin $k$ to destination $j$
and then construct an ``intensity'' of infected flux from origin $k$ to
destination $j$ at time $t$ by 
$
M_{k,t-h}^{out}P_{kj,t-h} (I_{kt}/N_{k})
$
with a lag $h>0$ in the mobility measure.
As there is a time lag between getting infected and showing symptoms,
our formulation takes into account that 
travelers from origin $k$ at time $t-h$ face case counts 
$I_{kt}$, which are recorded at $t$.
Given the ``intensity'' of daily  infected flux, 
we consider the external infection rate in region $j$ at time $t$ as follows:
\begin{align}
  \label{eq:lambda}
  \lambda_{jt} & =\frac{\theta_{t}}{N_{j}}\sum_{k\neq j}
                 M_{k,t-h}^{out}P_{kj,t-h} \frac{I_{kt}}{N_{k}},
\end{align}
The time-varying parameter $\theta_{t}$ reflects the strength of
external transmission and also normalizes the unit because the index $M_{kt}^{out}$
is a scaled measure. 
As in the specification for $\beta_{jt}$,
we allow $\theta_{t}$ to respond
to policy intervention gradually, i.e.,  
$
\log\theta_{t}=\log\theta_{t-1}+\rho X_{j,t-h},
$
where $\rho$ is a parameter.

\section{Data and Estimation Method}

\subsection{Data}

\begin{figure}[h]
  \centering
  \caption{Daily Outflux in 2020 and 2019}
  \includegraphics[scale=0.8]{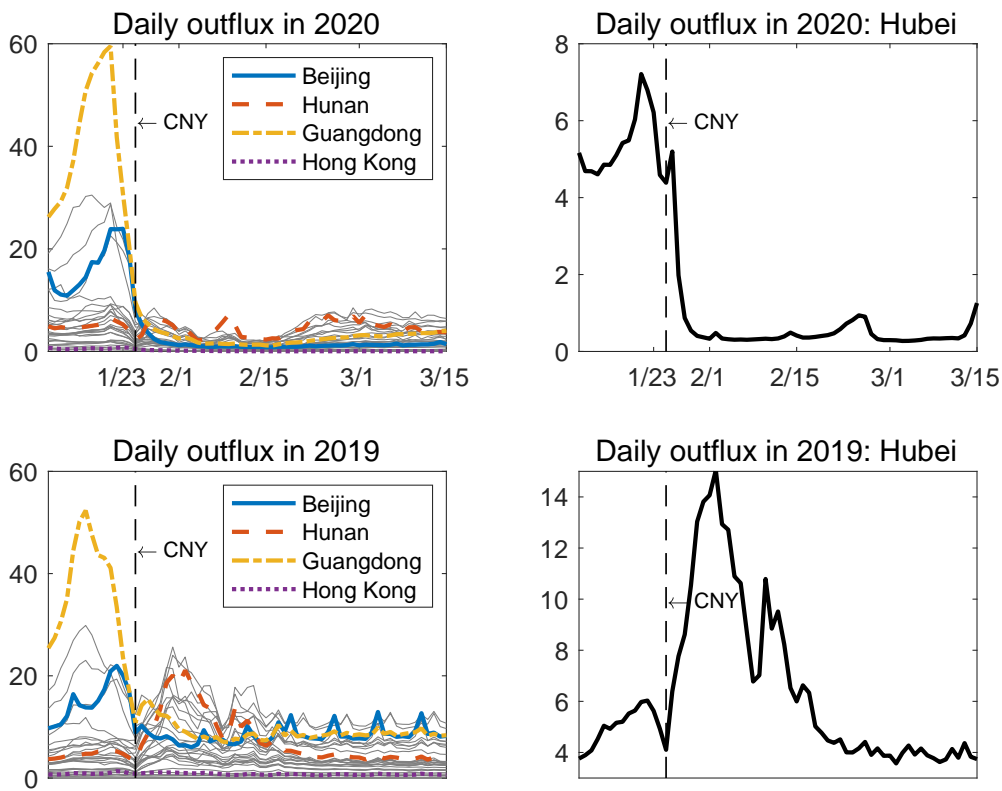}
  \label{fig:outflux}
  \begin{minipage}{17.5cm} \ \\
    \footnotesize
    \textit{Notes:}
    Two panels on the left column show daily outflux
    from all regions in 2020 and 2019.
    The ones on the right column show
    the outflux only in Hubei.
    In each panel, a dashed vertical line
    shows the date of the Chinese New Year (CNY) in 2019 and 2020.
  \end{minipage}
  \label{fig:rates}
\end{figure}

We use the daily data on COVID-19 infection and individuals' mobility 
from January 11 to March 15, 2020. 
The daily data of the infection, death, and recovery cases for each region 
are obtained from the National Health Commission of China
and its affiliates.\footnote{
  http://www.nhc.gov.cn/xcs/yqtb/list\_gzbd.shtml
} 
The human mobility data is obtained from
Baidu Migration \citep{Baidu}. The data provides a daily outflux index for each of
the 33 regions as well as the destinations of the outflux. 
For our counterfactual analysis, we use the mobility data set of 2019
from Baidu-Qianxi matched according to the Chinese New Year. 
The plots of the outflux in both 2020 and 2019 are shown in Figure \ref{fig:outflux}.
The outflux indices before the Chinese New Year in both 2019 and 2020
are dominated by provinces such as Guangdong, Zhejiang, and Beijing.
It is expected that most workers would be leaving these areas to return for their
home provinces for the holiday.
For Hubei, the outflux was moderate in both years. The outflux reduced to a negligible level at the time when the lockdown policy prevailed.

\subsection{Estimation Method}

We adopt a Bayesian framework for estimation. 
Given the information on infection, recovery and death cases,
we can estimate our model separately for the infection and the recovery and death.
In our model,
the number of recovered and death cases follows a multinomial distribution.
Thus, the likelihood of the parameters of recovery rate $\gamma$ and death rate $\delta$
has an analytic form.
We use a standard random-walk Metropolis sampler
with uninformative prior.

For the new case counts following 
the Poisson distribution, we simulate the posterior
distribution using the algorithm in \cite{ChibGreenbergWinkelmann1998}, which is based on data augmentation
and a Metropolis-Hastings-within-Gibbs sampler.
We divide the set of parameters into $J+2$ blocks:
$\{(\log\beta_{j,0},\alpha_{j})\}_{j=1}^{J}$, $(\mu,\Sigma)$,
and $(\log\theta_{0},\rho)$,
and then sample sequentially using their
conditional posteriors.
For each block of $\{(\log\beta_{j,0},\alpha_{j})\}_{j=1}^{J}$, 
we use a multivariate-$t$ proposal density whose mean and covariance
are computed from the mode and Hessian of the conditional posterior.
For $(\mu,\Sigma)$, we specify a Gaussian-inverse Wishart prior, $NIW(\mu^*,\kappa^*,\Lambda^*,\nu^*)$,
with $\mu^*=(-1,-0.1)'$, $\kappa^*=1$, $\Lambda^*=\text{diag}(1,0.05)$,
and $\nu^*=10$. This prior is weakly informative in $\mu$ and
moderately informative in the variance matrix $\Sigma$ . Lastly, the block
$(\log\theta_{0},\rho)$ is updated using a Gaussian prior
$N(\pi,\Omega)$
with $\pi=(0.1,-0.1)'$ and $\Omega=\text{diag}(0.1, 0.1)$.
As the posterior of $(\log\theta_{0},\rho)$ depends on all $J\times T$
observations, the contribution of the prior is minimal.


\section{Empirical Results}

This section first presents the estimation result for the heterogeneous internal infection rate and the effect of the regional intervention. We then compare the results of internal and external infection and provide additional findings based on a transmission network between regions.

\subsection{Transmission Parameter across Regions}

In Figure \ref{fig:rates}, we present the estimation result of internal transmission rates. Panel (a) of Figure \ref{fig:rates} shows the posterior means of
the initial transmission parameter, $\beta_{j,0}$.
The significant heterogeneity in the initial infection rate is evident here. Hubei has the highest value with a very tight posterior credible interval.
Panel (b) of Figure \ref{fig:rates} reports the transition of posterior means of the internal transmission rate $\beta_{j,t}$, which depicts the effects of policy intervention. The top-level health emergency response was activated for January 23-25, 2020, in all regions, except Xizang, which went into the state of emergency on January 30, 2020.
As in equation (\ref{eq:betat}), the number of new infections is shown to be affected by the policy implemented five days before. In Figure  \ref{fig:rates}, the effect of the intervention is evident but not immediate; it shows that for most regions, it took 4 to 7 days for $\beta_{jt}$ to decrease to half of its original value.
The posterior mean of recovery rate $\gamma$ is 4.15\% with a 95\% credible interval (4.11\%,4.18\%). The posterior mean of death rate $\delta$  and 0.213\%, with a 95\% credible interval (0.206\%,0.220\%).
\begin{figure}[!htb]
  \caption{Internal Transmission Rate}
  \subfloat[Initial Transmission Rate  ($\beta_{j,0}$)]{\includegraphics[width=8cm,height=12cm]{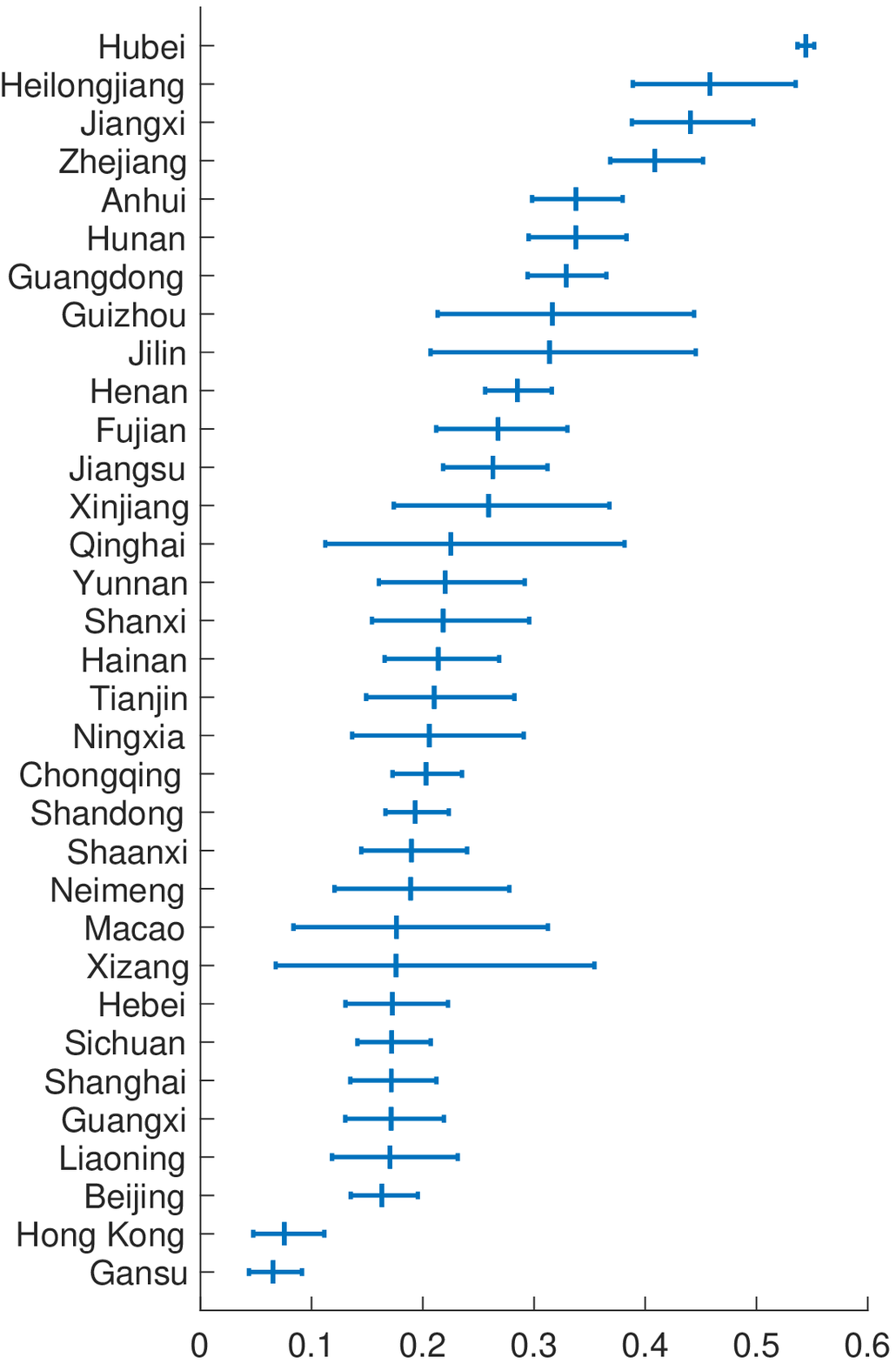}
  } \qquad{}
  \subfloat[Internal Transmission Rate ($\beta_{j,t}$) ]{\includegraphics[width=8cm,height=12cm]{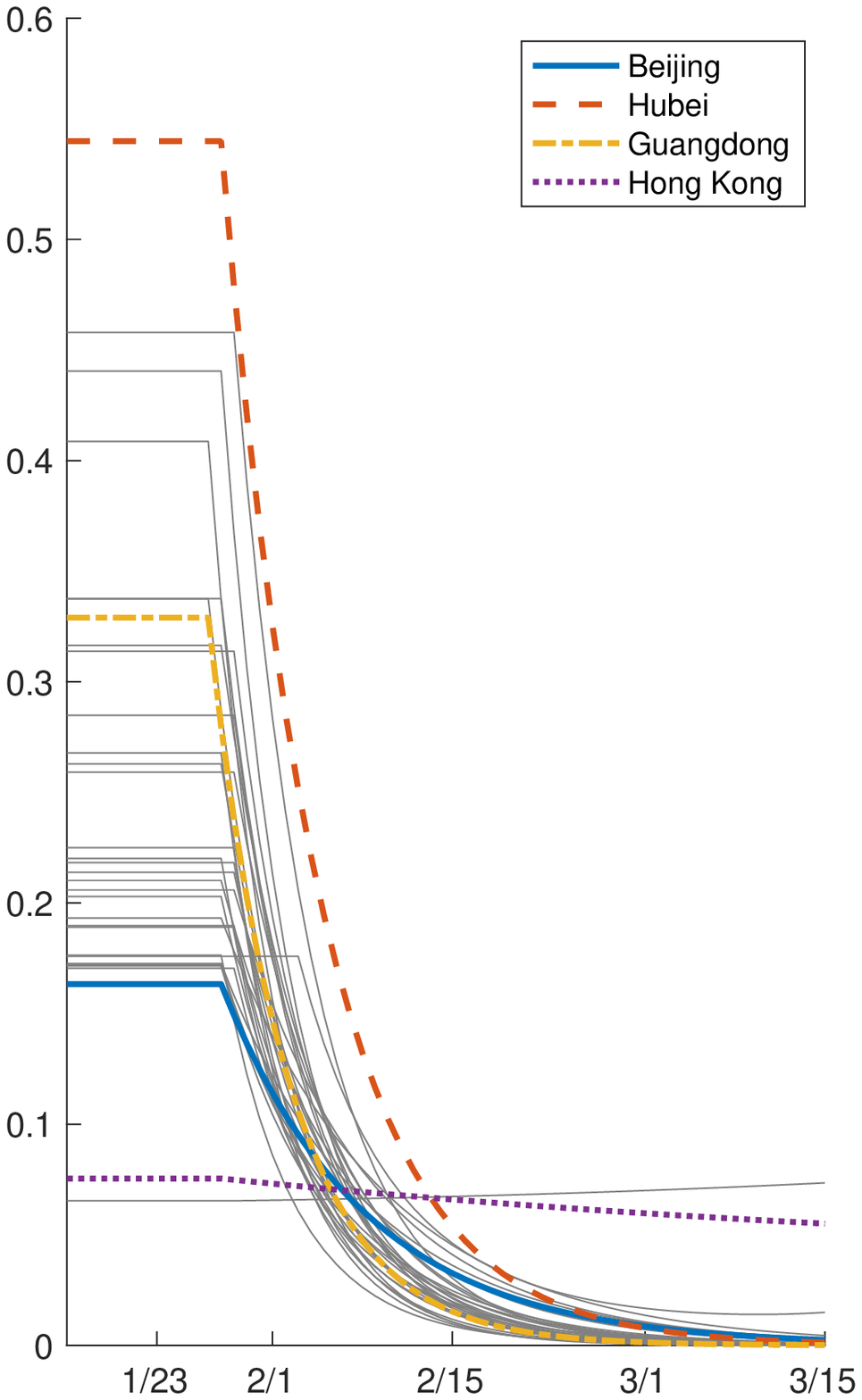}
  }

  \begin{minipage}{17.5cm} \ \\
    \footnotesize
    \textit{Notes:} 
    Panel (a) shows that 
    the posterior mean of the basic reproduction number
    for each region with 
    the line segment representing the 95\% posterior credible interval.
    Panel (b) reports the posterior mean of
    the effective reproduction number ($R_{t}$)
    across regions over time.
  \end{minipage}
  \label{fig:rates}
\end{figure}
\vspace{-0.6cm}

\subsection{Transmission Rates: External versus Internal}
\label{sec:transmissionrate}

\begin{figure}[!htb]
	\center
        \caption{The Number of Infected Individuals: External vs Internal Transmission}
	\includegraphics[scale=0.80]{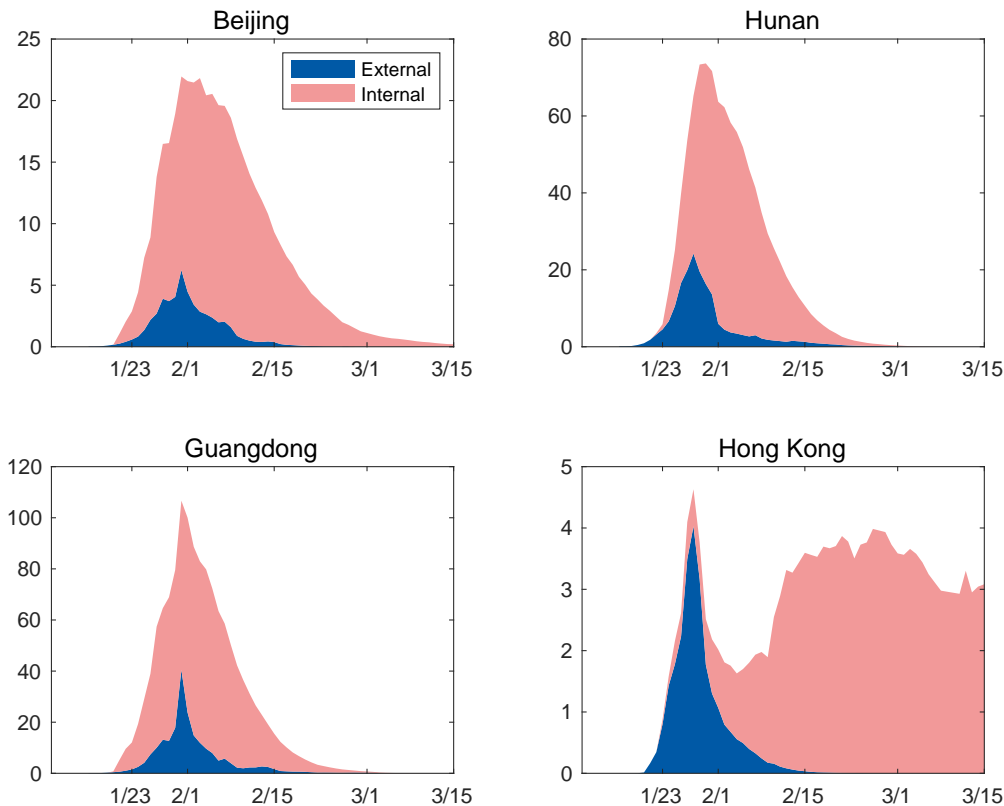}
	\label{fig:patterns} 

        \begin{minipage}{17.5cm} \ \\
          \footnotesize
          \textit{Notes:}
          The blue area represents the expected number of infections
          due to the external infection 
          and 
          the pink area represents the expected number of infections due to
          internal infection.
        \end{minipage}
\end{figure}

In combination with the lockdown policy, it is important to specifically study human mobility in the context of the COVID-19 pandemic. Our analysis decomposes the expected number of infections into infections resulting from internal and external transmission for all regions. Figure \ref{fig:patterns} presents results for four regions, each of which represents a different region, but all have similar characteristics. Namely, we consider megacities (Beijing), the neighboring regions of Hubei (Hunan), the secondary epicenters (Guangdong), and the special administrative regions outside mainland China (Hong Kong).

Internal transmission in Beijing and Guangdong follows a similar pattern with an exponential increase from the beginning of the outbreak, which dominates the external transmission influence. In these regions, there is an initial peak during the Chinese New Year (January 24 - February 2, 2020). This finding is empirical evidence that the pandemic had already expanded outside of Wuhan as early as late January 2020. By this time, other major cities can be considered to have been suffering from localized outbreaks already. On the other hand, the dominating form of transmission in Hunan is external until January 27. Similarly, in Hong Kong, external transmission dominates as the source of infection until February 5. Both internal and external transmission subsequently exhibit an exponential decrease due to unprecedented policy interventions, such as stay-at-home instructions and extended public holidays. The only exception to this observation is the evidence of internal transmission in Hong Kong, which still increased substantially following February 5 until it stabilized on February 15.

\begin{figure}[!htb]
  \center
  \caption{The Number of Infected Individuals due to External Transmission}
  \includegraphics[scale=0.8]{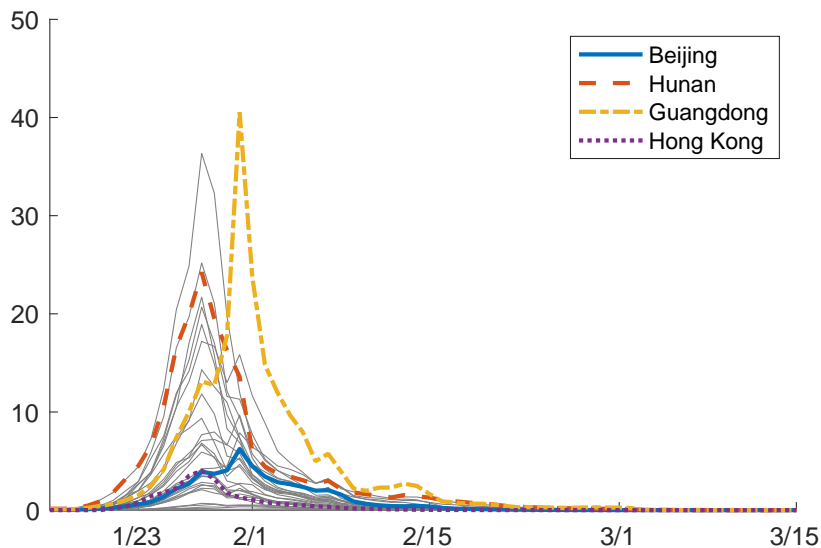}
\label{fig:external}
\begin{minipage}{17.5cm} \ \\
  \footnotesize
  \textit{Notes:}
  The expected number of infections due to external transmission
  is reported for all regions,
  and lines for the four representative cities are
  highlighted.
\end{minipage}
\end{figure}

Figure \ref{fig:external} presents the expected number of infections from external transmission in all regions. It is not surprising to see relatively high external infections in Hunan, an area that shares a border with Hubei. External transmissions peaked during the holiday periods across most regions and reduced to a negligible level by February 15. Guangdong shows the highest external infection rate, which also peaked later than most regions. This can be attributed to the large numbers of migrant workers returning from their hometown following the Chinese New Year.

\begin{figure}[!htb]
  \center
  \caption{External Transmission under Actual and Counterfactual Mobility }
\includegraphics[scale=0.80]{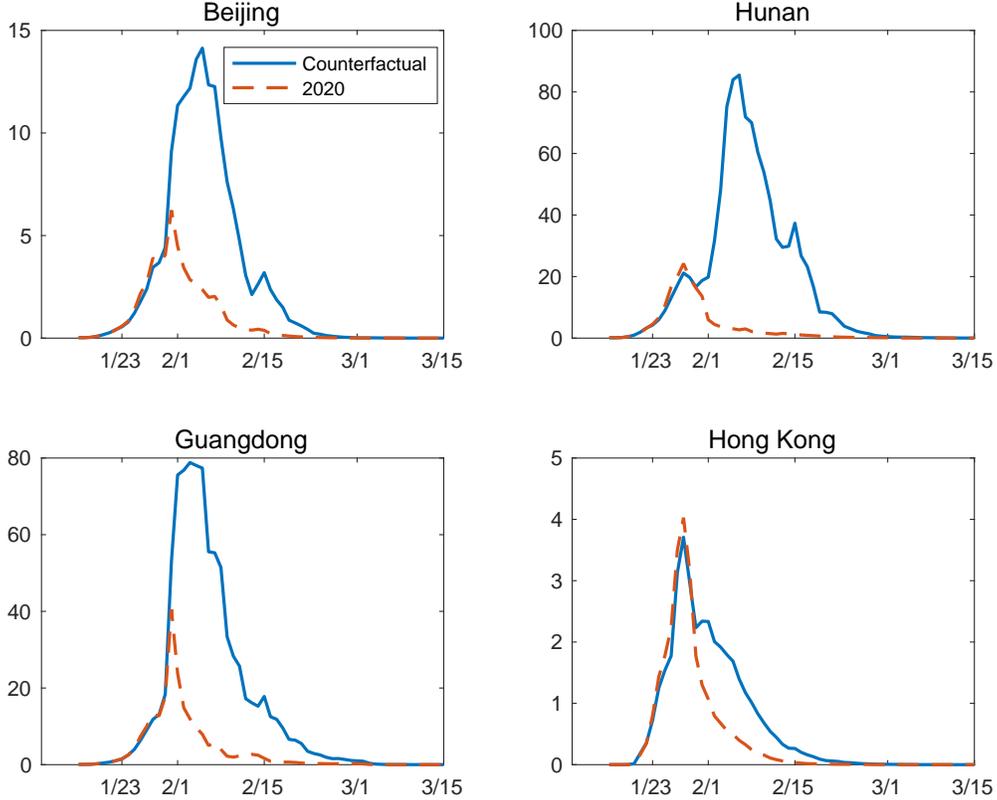} 
\label{fig:counter}
\begin{minipage}{17.5cm} \ \\
  \footnotesize
  \textit{Notes:}
  Panels (a)-(d) show the estimated number of
  infections due to the external transmission in four regions.
  The dashed (red) line shows the value based on mobility information in 2020,
  while 
  the solid (blue) line shows the value based on interactive fixed effects.
\end{minipage}
\end{figure}

To shed further light on these observations, we also conduct an analysis that applies 2019 mobility data.
Specifically, the lockdown effect is maintained but the 2020 mobility data is replaced with the 2019 data to isolate interactive fixed effects of mobility
across regions and time, following \cite{bai2009panel}.\footnote{
  For the counterfactual analysis,
  we first
  match the periods of 2019 and 2020 data 
  according to the Chinese New Year and 
  then extract interactive fixed effects
  from the 2019 data.
  This approach decomposes daily human mobility across regions
  into daily and regional factors 
  in a flexible yet parsimonious manner.
  More details are provided in the supplementary material.
}
The results are reported in Figure \ref{fig:counter}.
The first conclusion which can be drawn from this analysis is that the lockdown policy is effective. This is clear because the exponential decay of the external transmission is still evident across all 33 regions. However, if mobility is maintained at its usual level as indicated by the interactive fixed effects, the exponential decay would be delayed, resulting in much more significant external transmission rates across regions, except for Hong Kong. Hong Kong is an exceptional case because most of its external transmission originated from Guangdong (its only neighbor). Outflux from Guangdong in 2020 was significantly higher than its usual level in January. This led to a higher influx in Hong Kong during this time. Consequently, this abnormal influx in 2020 resulted in a high number of external transmissions in Hong Kong at the early stage of the outbreak.

\subsection{Transmission Network} 

The transmission network between the regions of China is observed to evolve on each day of the pandemic.
\citet{kraemer2020effect} focuses on the transmission from Wuhan to the rest of China and they conclude that the propagation of COVID-19  in China during the early stage of the outbreak was mostly explained by human mobility originated from Wuhan.
However, the authors did not consider the mobility network among the rest of China's geography, and thus, the scope of analysis of the transmission channels is limited. The main advantage of the model developed in this study is that it enables the transmission network to be analyzed on a more granular level. This means that the sources of external transmission and their respective intensities can be identified.
Specifically, based on (\ref{eq:lambda}), we can obtain the rate of external transmission from region $k$  to region $j$, 
\begin{eqnarray*}
  A_{jkt}:=\frac{\theta_{t}}{N_{t}} M_{k,t-h}^{out}P_{kj,t-h} \frac{I_{kt}}{N_{k}}S_{jt}  .
\end{eqnarray*}
Following the literature on network theory
\citep[e.g.,][]{aldous2003graphs}, 
we can interpret the square matrix consisting of $A_{jkt}$ for $j, k = 1, \dots, 33$
as an adjacency matrix of a directed graph
with weighted directions $A_{jkt}$ from $k$ to $j$ at time $t$.
The sum $\sum_{j \neq k}A_{jkt}$ represents the diseases transmissions which originated from region $k$ and moved to the other regions at time $t$.

\begin{figure}[!htb]
  \center 
  \caption{Origins of Transmission: the effective infected outflux from each region over time}
  \includegraphics[scale=0.6]{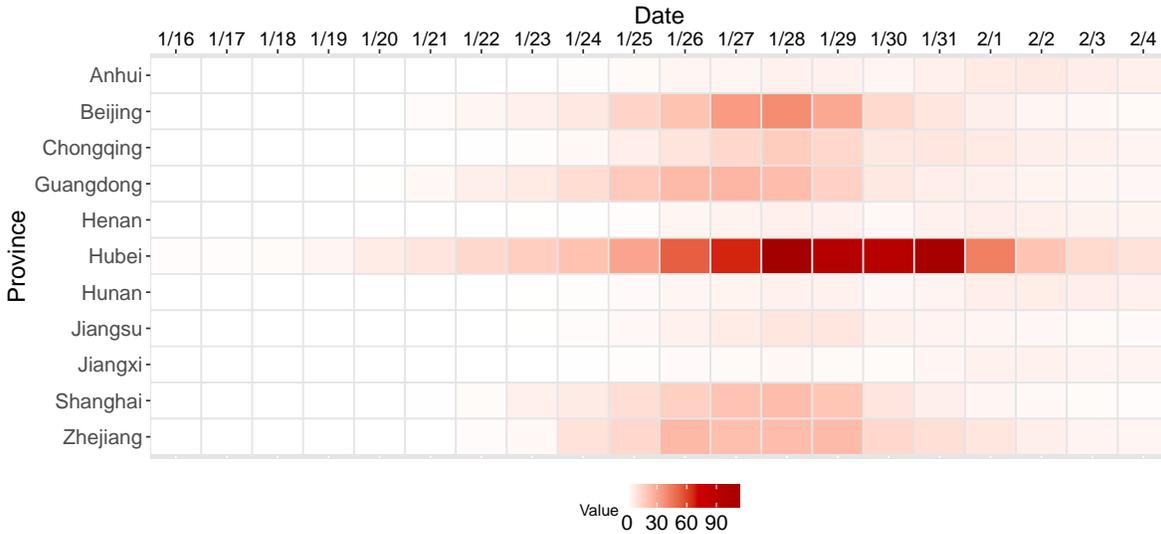}
  \label{fig:epicenter}	
\begin{minipage}{17.5cm} \ \\
  \footnotesize
  \textit{Notes:}
  The horizontal axis shows days from January 16 to February 4
  and 
  the vertical axis shows ten regions, which
  are the origins of the ten highest external daily transmission
  to the other regions.
  The heatmap reports values of $\sum_{j \neq k}A_{jkt}$ for each origin $k$.

\end{minipage}
\end{figure}

Figure \ref{fig:epicenter} presents
the heatmap of $\sum_{j \neq k}A_{jkt}$,
thus clarifying the top 10 most influential regions, that is, the regions which were the source of the most transmissions, over time.
Hubei stands out as the primary exporter of the infection during the Chinese New Year holidays, though results show that secondary epicenters, such as Beijing, Guangdong, and Shanghai, started being a significant source of transmission from around January 22. The outflux from epicenters, including the primary one, Hubei, gradually diminished following the enactment of policy interventions.

\begin{figure}[!htb]
  \center
  \caption{Transmission Network on January 27, 2020}
  \includegraphics[trim=1.7cm 0 1.7cm 1.3cm,clip,width=\linewidth]{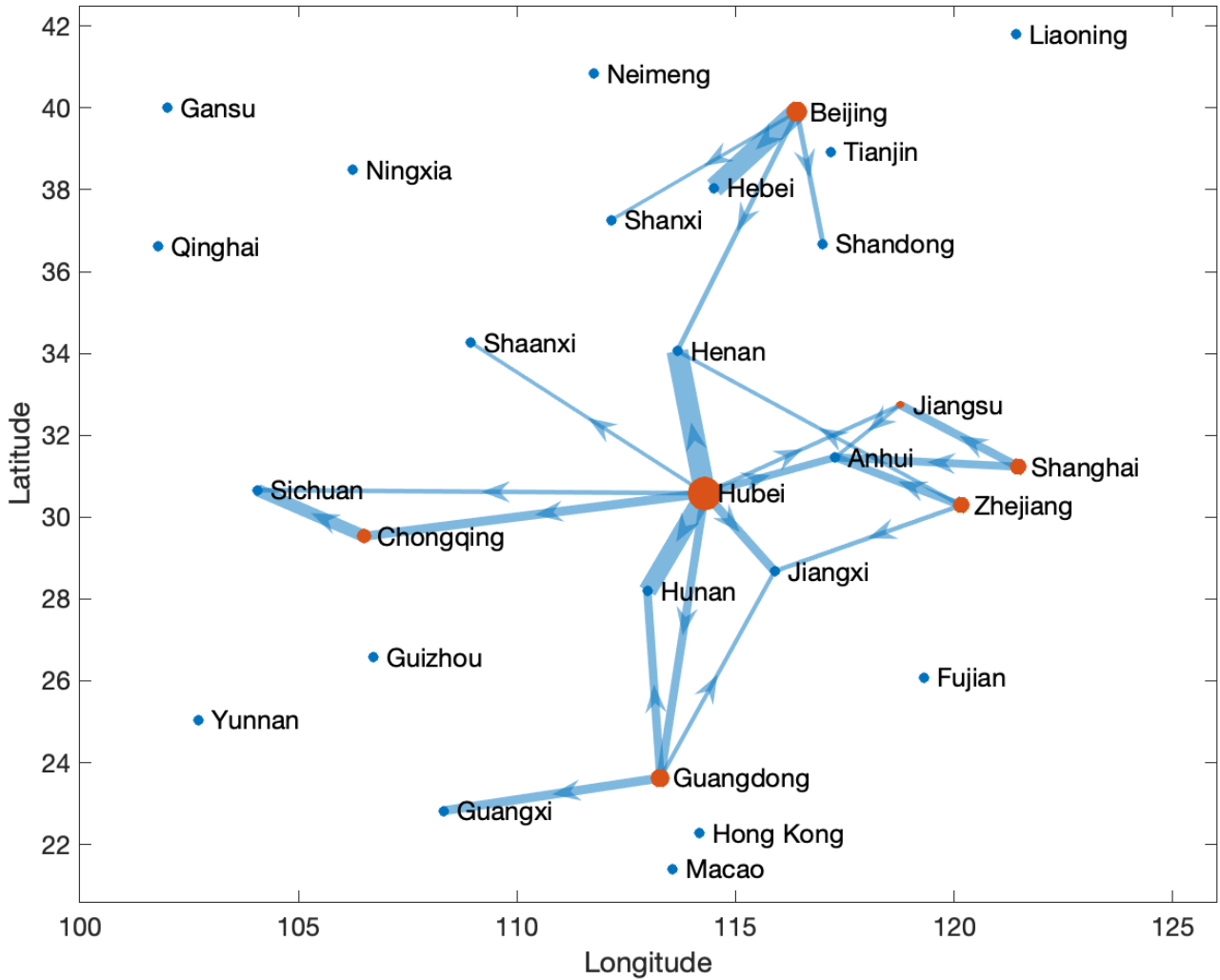}
\begin{minipage}{17.5cm} \ \\
  \footnotesize 
  \textit{Notes:}
  Regions are located geographically. The
    arrow indicates the direction of transmission, the lines display external transmissions that are greater than two. The line widths are proportional to the external transmission in the indicated direction and the size of the nodes are proportional to the total export from a region.
\end{minipage}
\label{fig:network}
\end{figure}

To examine the transmission network more closely, we present a section of the transmission network on January 27, 2020, in Figure \ref{fig:network}. All the regions in the network are depicted in the figure according to their geographic location. The arrows reflect transmission directions. The lines display transmissions with $A_{kjt}$ greater than two, whereby the line width is proportional to the transmission strength. Figure \ref{fig:network} shows that Hubei is the primary epicenter, particularly for geographically proximate provinces (e.g., Henan and Hunan), but also that secondary epicenters such as Beijing, Guangdong, and Shanghai, have already developed on this date. The dynamic migration between the secondary epicenters---which are cultural and economic centers---and the rest of China accelerated the propagation of the disease. Regions such as Shandong and Guangxi are only influenced by secondary epicenters, whereas regions such as Sichuan evidence disease transmission originating from both the primary epicenter and the secondary epicenters. 

\section{Conclusion}

We analyze the propagation of COVID-19 among 33 provinces and special administrative regions in China. We develop a spatial model to estimate the transmission network and evaluate the effect of the policy interventions of lockdown and mobility restrictions on the disease spread. Our empirical results suggest that secondary epicenters developed at a very early stage of the epidemic and that mobility restriction across provinces prevented further spread of the disease. Thus, the observed pandemic propagation in China could be re-contained to localized outbreaks. Community transmission was observed to be the primary source of infection, and regional policy intervention stemmed the spread of the disease. Our empirical findings suggest that the coordination of central and local government policies is essential to suppress the spread of infectious diseases.

\clearpage
\setstretch{0.15} \setlength{\bibsep}{6pt} 
\bibliographystyle{ecta}
\bibliography{nCov}

\end{document}